\documentclass[runningheads]{svmult}

\usepackage{epsfig}

\usepackage{makeidx}   
\usepackage{graphicx}  
\usepackage{subeqnar}  
\usepackage{multicol}  
\usepackage{physprbb}  
\makeindex             

\begin{document}
\title*{Dynamics and Thermodynamics of Systems with Long Range
   Interactions: an Introduction}

\toctitle{Dynamics and Thermodynamics of Systems with Long Range
   Interactions: an Introduction}

\titlerunning{Dynamics and thermodynamics of systems with long range
   interactions}

\author{Thierry Dauxois\inst{1}
\and Stefano Ruffo\inst{2}
\and Ennio Arimondo\inst{3}
\and Martin Wilkens\inst{4}}

\authorrunning{T. Dauxois et al.}
\institute{ Laboratoire de Physique, UMR CNRS 5672, ENS Lyon, 46,
all{\'e}e d'Italie, F-69007 Lyon, France  \and Dipartimento di 
Energetica "S. Stecco", Universit{\`a} di
Firenze, via S. Marta, 3, I-50139 Firenze, Italy \and Dipartimento di Fisica,
Universit{\`a} degli Studi di Pisa Via F. Buonarroti, 2 I-56127, Pisa,
Italy \and
Universit{\"a}t Potsdam, Institut fuer Physik Am Neuen Palais 10,
14469 Potsdam, Germany }

\maketitle              

\begin{abstract}
We review  theoretical results obtained recently in the
framework of statistical mechanics to study systems with long range
forces. This fundamental and methodological study leads us to consider
the different domains of applications in a trans-disciplinary
perspective (astrophysics, nuclear physics, plasmas physics,
  metallic clusters, hydrodynamics,...)
with a special emphasis on Bose-Einstein condensates.

\end{abstract}

\section{Introduction}

Properties of systems with long range interactions are to a large
extent only poorly understood although they concern a wide range of
problems in physics. Recently, the disclosure of new methodologies to
approach the study of these systems has revealed its importance also
in a trans-disciplinary perspective (astrophysics\index{astrophysics},
nuclear physics\index{nuclear physics}, plasmas physics,
Bose-Einstein condensates\index{Bose!Einstein condensation}, atomic
clusters\index{atomic!clusters}, hydrodynamics,...). The main
challenge is represented by the construction of a thermodynamic
treatment of systems with long range forces and by the understanding
of analogies and differences among the numerous domains of
applications.

Some promising results in this direction have been recently obtained
in the attempt of combining tools developed in the framework of
standard statistical mechanics with concepts and methods of dynamical
systems\index{dynamical systems}.  Particularly arduous, but very
exciting, is the understanding of phase transitions (negative specific
heat, non extensive thermodynamics,...)  for such systems as well as
all the aspects related to non-equilibrium phenomena and their
description in terms of dynamical concepts (self-consistent
chaos\index{self!consistent chaos}, slow relaxation, formation and
role of structures\index{coherent!structures},...).

Finally this fundamental and methodological study should help us to
detect the depth and the origin of the analogies found in the
different domains mentioned above or on the contrary emphasize their
specificities.  In particular, we would like to put a special emphasis
on Bose Einstein Condensation (BEC) which could be the main field of
applications, since experiments and theoretical ideas have reached an
impressive quality in the last decade. In that domain, many
inequivalences between ensembles have been reported and should be
clarified. Moreover, long range interactions in BEC have opened very
exciting new perspectives to consider BEC as a model for other
systems.

\section{Why systems with long range interactions are important ?}

\subsection{The problem of additivity\index{additivity}
}
\label{Theproblemofadditivity}

The methods to describe a given system of $N$ particles
interacting via a gravitational potential in  $1/r$ are
dramatically dependent on the value $N$. If Newton showed the
exact solution for $N=2$, and one can expect to get a numerical
solution in the range $N=3-10^3$, the results are clearly out of
reach for a larger number of particles. In addition,
it is clear that the knowledge of the evolution of the different
trajectories is completely useless, since it is well known that
these systems are chaotic as soon as $N$ is greater than two.
Therefore, one needs to get a {\em statistical} analysis, in order
to get insights in the thermodynamical properties~\cite{Padmanabhan} of the system
under study.

However, such statistical study leads immediately to unexpected
behaviors for physicists used to neutral gases, plasmas or atomic
lattices. The underlying reason is directly related to the long
range of the interaction, and more precisely to the non additivity
of the system.

To avoid misunderstandings, let us first clarify the definition of
{\em extensivity}\index{extensivity} with respect to {\em additivity}.
A thermodynamic variable, like the energy or the entropy, would be
extensive, if it is proportional to the number of elements, once the
intensive variables are kept constant. To be more precise, let us
consider the following Hamiltonian
\begin{eqnarray}
H=-\frac{J}{N}\left(\sum_{i=1}^N S_i\right)^2\quad,\label{hamiladditi}
\end{eqnarray}
where the spins $S_i=\pm1$ sit on a one-dimensional
lattice labeled by $i=1,\ldots,N$. Without the $1/N$ prefactor  such a
Hamiltonian would  have an ill defined thermodynamic limit. This
is correctly restored by applying the Kac
prescription~\cite{KacUhlenbeck}\index{Kac prescription}, 
within which the potential is
rescaled by an appropriate volume dependent factor, here proportional
to $N$ since the lattice is one-dimensional: such a
Hamiltonian is therefore extensive. Let us note in passing that
this regularization is not always accepted. In cases with a
kinetic term, such a regularization corresponds to a
renormalization of the time scale.

\begin{figure}[ht]
\begin{center}
\includegraphics[width=.3\textwidth]{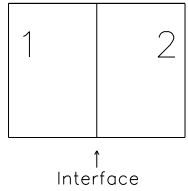}
\end{center}
\caption[]{Schematic picture of a system separated in two equal
parts.} \label{additivite}
\end{figure}

On the contrary, this Hamiltonian is not additive. Indeed, let us
divide a system schematically pictured in Fig.~\ref{additivite}, in
two equal parts. In addition, one considers the particular case with
all spins in the left part are equal to 1, whereas all spin in the
right part are equal to -1. It is clear that the energy of the two
different parts, will be
$E_1=E_2=-\frac{J}{N}\left(\frac{N}{2}\right)^2=-\frac{JN}{4}$.
However, if one computes the total energy of the system, one gets
$E=-\frac{J}{N}\left(\frac{N}{2}-\frac{N}{2}\right)^2=0$. It is
therefore clear that such a system is not additive, since one cannot
considers that $E_1+E_2= E$, even approximately. The energy of the
interface, usually neglected, is clearly of the order of the energies
of the two different parts: the system is not additive. The underlying
reason is that Hamiltonian~(\ref{hamiladditi}) is long (strictly
speaking infinite) range\index{mean field!models}, 
since every spin interact with all others:
moreover, as the interaction is not dependent on the distance between
spins, this is a mean-field model. This example is further elaborated
in~\cite{BMRdd}.

This non additivity has strong consequences in the construction of the
canonical ensemble\index{canonical!ensemble}. Once the microcanonical
ensemble\index{microcanonical!ensemble} has been defined, the usual
construction of this ensemble is usually taught as follows. The
probability that system $1$ has an energy $E_1$ and system $2$ has an
energy $E_2$ is proportional to $\Omega_1(E_1)\; \Omega_2(E_2)\;dE_1$, where the
number of states of a system with a given energy $E$, $\Omega(E)$, is
related to the entropy via the classical Boltzmann formula $S(E)=\ln
E$ (we omit the $k_B$ factor for the sake of simplicity). Using the
additivity of the energy, and considering the case where system $1$ is
much smaller than system $2$, one can expand the term $S_2(E-E_1)$, as 
shown in the following different steps
\begin{eqnarray}\label{derivdistcanon}
\Omega_1(E_1)\; \Omega_2(E_2)\;dE_1&=&\Omega_1(E_1)\;\Omega_2(E-E_1)\;dE_1\\
&=&\displaystyle\Omega_1(E_1)\; e^{\displaystyle S_2(E-E_1)}\;dE_1\\
&=&\Omega_1(E_1)\;e^{ \displaystyle(S_2(E)-E_1\frac{\partial 
S_2}{\partial E}_{|E}+...)}\;dE_1\\
&\propto&\Omega_1(E_1)\;e^{ \displaystyle-\beta E_1}\;dE_1\quad,
\end{eqnarray}
where $\beta=\frac{\partial S_2}{\partial E}_{|E} $. One ends up with the usual
canonical distribution\index{canonical!partition function}. It is
clear that additivity\index{additivity} is crucial to
justify~(\ref{derivdistcanon}), which means that non additive system
will have a very peculiar behavior if there are in contact with a
thermal reservoir. This is one of the topic discussed in this paper,
and in numerous contributions in this book.

\subsection{Definition of long range systems}

\begin{figure}
\begin{center}
\includegraphics[width=.3\textwidth]{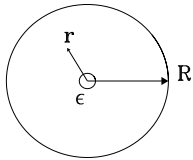}
\end{center}
\caption[]{Schematic picture of a particle interacting with all
particles located in a homogeneous sphere of radius $R$, except
the closest ones located in the sphere of radius~$\varepsilon$.}
\label{sphere}
\end{figure}

To define now systems with long range interactions, let us
consider the potential energy for a given particle, situated in
the center of a homogeneous sphere with radius $R$. We will omit
at this stage the interactions of the particles situated in the
neighborhood of the particles by forgetting particles located in
the sphere of radius $\varepsilon$, where $\varepsilon\ll R$ as
shown in Fig.~\ref{sphere}. The reason will be explained in the
following subsection.

If one considers that particles interact via a potential energy
proportional to ${1}/{r^\alpha}$, where $\alpha$ is the
key-parameter defining the range of interaction, we obtain in the
three dimensional space
\begin{eqnarray}
U=\int_\varepsilon^R4\pi r^2dr\;
\rho\;\frac{1}{r^\alpha}=4\pi\rho\int_\varepsilon^Rr^{2-\alpha}dr\propto\left[r^{3-\alpha}\right]_\varepsilon^R
\end{eqnarray}
where $\rho$ is the particle density. When increasing the radius $R$, the
contribution due to the surface of the sphere, $R^{3-\alpha}$, could be
neglected when $\alpha>3$, but diverges if $\alpha<3$. In the latter case,
surface effects are important and therefore additivity is not
fulfilled.

If one generalizes this definition to long range systems in $d$
dimensions, one easily shows that energy will not be additive if
the potential energy behaves at long distance as
\begin{eqnarray}
V(r)\sim \frac{1}{r^\alpha}\quad \mbox{with}\quad
\frac{\alpha}{d}<1\quad.
\end{eqnarray}
Mean-field models\index{mean field!models}
 as Hamiltonian~(\ref{hamiladditi}) corresponds
to the  value $\alpha=0$, since the interaction does not
depend on the distance. There are therefore not additive as shown
in section~(\ref{Theproblemofadditivity}). J. Barr{\'e} {\it et al}
consider~\cite{BMRdd} such a mean field model: the
Blume-Emery-Griffith (BEG) model with infinite range interactions.
The gravitational problem, which is at the origin of this study, and
corresponds to $\alpha=1$ in three dimensions, clearly belongs to
this category, but presents also additional difficulties.

\subsection{Difficulties with the gravitational problem\index{astrophysics}
\index{gravitation}
}

This problem is particularly tedious because, in addition to the non
additivity due to long range character, such a system needs a
careful regularization at short distances to avoid
collapse. To be more specific, let us consider the configurational
partition function\index{canonical!partition function}
 of a system of $N$ particles
\begin{equation}\label{cutoffgravit}
  Z_U=\int_{V} \displaystyle d^{3N}\overrightarrow{r_i}\ e^{-\beta
U(\overrightarrow{r_i})}\quad,
\end{equation}
where one notes $U(\overrightarrow{r_i})$ the gravitational
potential energy, $\beta$ the inverse of the temperature and $V$
the volume of the system.
\begin{figure}
\begin{center}
\includegraphics[width=.5\textwidth]{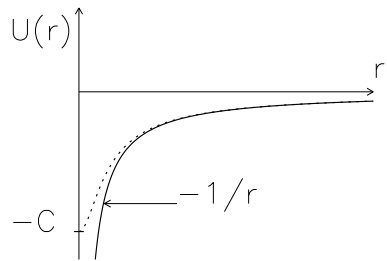}
\end{center}
\caption[]{The gravitational potential energy as a function of the
distance $r$ is represented by the solid curve, whereas the dotted
one shows the regularized potential energy to avoid gravitational
collapse.} \label{courtedistance}
\end{figure}
From the shape of the potential energy depicted in
Fig.~\ref{courtedistance}, one clearly see that $Z_U$ will diverge if
all particles collapse towards the same point. This difficulty arises
because the potential energy is {\em not bounded from below} as for a
Lennard-Jones\index{Lennard-Jones potential} or a Morse potential.
This effect is of course physically forbidden by the Pauli principle.
However, to avoid the use of Quantum Mechanics, the usual trick is to
introduce an add-hoc cut-off. The potential is therefore "regularized"
by introducing the value $-C$, shown in Fig.~\ref{courtedistance}.
Thus, the inequality $U(\overrightarrow{r_i})\geq-C$ allows easily to
find a {finite} upper bound for the configurational partition function
$Z_U\leq V\;e^{\beta C}$, where $V$ is the volume of the system.

However, there is a third difficulty in the case of gravitational
interaction: the system is open, i.e. without boundary, strictly
speaking. In the microcanonical ensemble
\index{microcanonical!ensemble}, the number of states
\begin{eqnarray}
   \Omega(E)=\int\prod_idp_i\int \prod_idq_i\;
   \delta\left(E-H\left(q_i,p_i\right)\right)
\end{eqnarray}
will diverge if the system is not confined. This divergence is
actually not restricted to the gravitational interaction but would
also occur if one considers a perfect gas in an infinite volume.
However, one considers of course always a gas in a finite domain, i.e.
in a finite volume. This is not any more possible for the
gravitational interaction where the system is clearly infinite.

Despite these additional difficulties, the
astrophysics\index{astrophysics} \index{gravitation} community has
obtained an impressive quantity of results in this domain. Thanu
Padmanabhan~\cite{paddy}  describes several remarkable features,
both for isolated gravitating systems as well as for systems
undergoing nonlinear clustering in an expanding background cosmology.
The emphasis is on general results and he brings out the
inter-relationships of this subject with topics in fluid mechanics,
condensed matter and renormalization group theory.

Similarly, Pierre-Henri Chavanis~\cite{chavanis} presents how the
structure\index{coherent!structures} and the organization of stellar
systems (globular clusters, elliptical galaxies,...) in astrophysics
can be understood in terms of a statistical
mechanics\index{statistical mechanics} for a system of particles in
gravitational interaction. Finally, Eddie Cohen and Iaroslav
Ispolatov~\cite{cohen} consider the related gravitational-like
collapse of particles with an attractive $1/r^\alpha$ potential.  Using
mean field continuous integral equation, they determine the
saddle-point density profile that extremizes the entropy functional.
For all $0 < \alpha < d=3$, a critical energy is determined below which the
entropy of the system exhibits a discontinuous jump.

\subsection{Applications to large systems}

However a growing scientific community has recently begun to
tackle this problem with different viewpoints. One of the fascinating
aspects of this problem is that, in addition to gravitating
systems, it concerns a large variety of systems that we would like
to discuss briefly in the following section.

\subsubsection{Plasmas\index{plasma!physics}}: 
Rarefied plasmas share many properties with collisionless stellar
systems, and in particular that the mean field \index{mean field!models}
 of the system is more important than the fields of
individual nearby particles.  Here again, the Coulomb
force\index{Coulomb interaction} is of long range character. However,
there is a fundamental difference between plasmas and gravitation.
Plasmas have both positive and negative charges, so that they are
neutral on large scales and can form static homogeneous equilibria; on
the contrary, gravitating systems can never form static homogeneous
equilibria. This so-called Debye screening explains why many
techniques of plasma physics can not be transferred immediately.
Elskens~\cite{elskens} and Diego Del Castillo Negrete~\cite{diego}
present some of their results in the framework of plasma physics.

\subsubsection{2D Hydrodynamics}:\index{hydrodynamics}
Two-dimensional incompressible hydrodynamics is another important
case where the interaction is long range. Indeed, the
streamfunction $\psi$ is related to the modulus of the vorticity
$\omega$, via the Poisson equation $\Delta\psi=\omega$. Using the
Green's function technique, one easily finds that the solution is
\begin{equation}
   \psi( \overrightarrow{r})=-\frac{1}{2\pi}\int_D d^2
\overrightarrow{r}'\;
\omega(\overrightarrow{r}')\;G\left(\overrightarrow{r}-\overrightarrow{r}'\right)\quad,
\end{equation}
where $G\left(\overrightarrow{r}-\overrightarrow{r}'\right)$ depends
on $D$, but $G\left(\overrightarrow{r}\right) \sim|\ln
\overrightarrow{r}|$, when $\overrightarrow{r}\to 0$. The kinetic energy being conserved by the Euler
equation (dissipativeless), it is straightforward to compute it on the
domain $D$, with boundary~$\partial D$,
\begin{eqnarray}\label{}
E&=&\int_D d^2 \overrightarrow{r}
\frac{1}{2}\left(\nabla\psi\right)^2\\&=&\oint_{\partial D}
\overrightarrow{n}dl\ \psi\nabla\psi+\frac{1}{2}\int_D d^2
\overrightarrow{r}\
\omega(\overrightarrow{r})\psi(\overrightarrow{r})\\
&=&-\frac{1}{4\pi}\int\int_D d^2\overrightarrow{r} d^2
\overrightarrow{r}'\;
\omega(\overrightarrow{r}')\omega(\overrightarrow{r})\ln|\overrightarrow{r}-
\overrightarrow{r}'|
\end{eqnarray}
since $\psi=0$ on $\partial D$.  This emphasizes that one gets a logarithmic
interaction. The analogy is even more clear if one approximates the
vorticity field  by point vortices
$\omega(\overrightarrow{r})=\sum_i\Gamma_i\delta(\overrightarrow{r}-\overrightarrow{r}_i)$,
located at $\overrightarrow{r}_i$, with a given
circulation~$\Gamma_i$. The energy of the system reads now
\begin{equation}
E=\frac{1}{2}\sum_{i\neq
j}\Gamma_i\Gamma_j\ln|\overrightarrow{r}_i-\overrightarrow{r}'_j|\quad.
\end{equation}
The interaction among vortices has a
logarithmic character, which corresponds to $\alpha=0$.

Pierre-Henri Chavanis~\cite{chavanis}  studies carefully the
analogy between the statistical of large-scale vortices in
two-dimensional turbulence and self-gravitating systems. This
analogy concerns not only the equilibrium states, i.e. the
formation of large-scale structures, but also the relaxation
towards equilibrium and the statistics of fluctuations. Diego Del
Castillo Negrete~\cite{diego} discusses also his results in the
framework of hydrodynamics.

\subsubsection{Dipolar interactions}\index{dipolar interaction}

Dielectrics and diamagnets in an external electric or magnetic field
exhibit a shape dependent thermodynamic limit \index{thermodynamic limit}
 \cite{Landau}. This is due to the marginal decay of the
potential energy $\alpha=d=3$ for systems of dipoles. There is some
approach to the solution of this problem only in zero field and in the
absence of spontaneous ferromagnetism \cite{Griffiths}.  This is a
border case for the long-range interactions, but it deserves a special
attention.

\subsubsection{Fracture}\index{fracture}

Let us examine analytical solutions for the plane stress and
displacement fields around the tip of a slit-like plane crack in an
ideal Hookean continuum solid. The classic approach to any linear
elasticity problem of this sort involves the search for a suitable
``stress function'' that satisfies the so-called biharmonic equation
$\nabla^2(\nabla^2\psi)=0$ where $\psi$ is the Airy stress function, in accordance
with appropriate boundary conditions. The deformation energy density
is then defined as $U\propto\sigma \varepsilon$ where $\sigma$ is the fracture stress field
around the tip, whereas $\varepsilon$ is the deformation field. Considering a
crack-width $a$ in a two-dimensional material and using the exact
Muskhelishvili's solution~\cite{muskhelishvili}, one otbains the
elastic potential energy due to the crack
\begin{equation}
U\simeq\frac{\sigma_\infty^2(1-\nu)}{2E}\;\frac{a^2}{r^2}\quad,
\end{equation}
where $E$ is the Young modulus, $\sigma_\infty$ the
stress field at infinity, $\nu$ the Poisson coefficient and $r$ the
distance to the tip: the elasticity equation in the bulk of solids
leads therefore, again, to a border case for the long-range
interactions since $U\sim 1/r^2$ in two dimensions. It appears that,
despite of its engineering applications, the dynamics of this non
conservative system has been very little studied, presumably because
of its long range character.  In addition, in such a two dimensional
material, the presence of several fractures could exhibit very
interesting screening effects.

\begin{table}
\caption{Table listing different applications where systems are
governed by long range interactions. {\em Large systems} where the
interactions is truly long range and {\em small systems} where the range
of the interactions is of the order of the size of the system are
separated.}
\begin{center}
\renewcommand{\arraystretch}{1.4}
\setlength\tabcolsep{5pt}
\begin{tabular}{llll}
\hline\noalign{\smallskip} Interactions & $\alpha$
&
  $\alpha/d$ & Comments\\
\noalign{\smallskip} \hline \noalign{\smallskip}
    \mbox{Large systems} &  &  &\\
\noalign{\smallskip} \hline \noalign{\smallskip}
   \mbox{Gravity\index{gravitation}
} & 1 & 1/3 &  \mbox{Long range}\\
   \mbox{Coulomb\index{Coulomb interaction}
}& 1 & 1/3 & \mbox{Long range with Debye screening}\\
   \mbox{Dipole\index{dipolar interaction}
}& 3 & 1 & \mbox{Limiting value}\\
\mbox{2D Hydrodynamics\index{hydrodynamics}
} & 0 & 0 & \mbox{Logarithmic interactions}\\
\mbox{Fracture\index{fracture}
} & 1 & 1 & \mbox{Stress field around the tip}\\
\noalign{\smallskip} \hline \noalign{\smallskip}
  \mbox{Small systems} &  &  & \\
\noalign{\smallskip} \hline \noalign{\smallskip}
  \mbox{atomic and molecular clusters\index{atomic!clusters}
} &  &  &\\
  nuclei\index{nuclear physics}
 &  &  & \\
  \mbox{BE Condensation\index{Bose!Einstein condensation}
} &  &  &\\
  \noalign{\smallskip} \hline \noalign{\smallskip}
\end{tabular}
\end{center}
\label{apptab1b}
\end{table}

\subsection{Applications to small systems}

In addition to large systems where the interactions is truly long
range, one should consider small systems where the range of the
interactions is of the order of the size of the system. In these
cases, the system would not be additive either and many similarities
will be encountered. Phase transitions\index{phase!transition} are
universal properties of interacting matter which have been widely
studied in the thermodynamic limit of {\it infinite} systems.
However, in many physical situations this limit is not accessible and
phase transitions should be considered from a more general point of
view.  This is for example the case for some microscopic or mesoscopic
systems: atomic clusters\index{atomic!clusters} can melt, small drops
of quantum fluids may undergo a Bose-Einstein condensation or a
super-fluid phase transition, dense hadronic matter 
\index{nuclear physics} is predicted to merge in a quark and gluon plasma phase
while nuclei are expected to exhibit a liquid-gas phase transition.
For all these systems the experimental issue is how to characterize a
phase transition in a {\it finite} system.

Philippe Chomaz and Francesca Gulminelli~\cite{chomazdd}, discusses
results from nuclear physics as well as from clusters physics. In
particular, they introduce a possible definition of first order phase
transitions in finite systems based on topology anomalies
\cite{Dieter} of the event distribution in the space of observations.
This generalizes the definitions based on the curvature anomalies of
thermodynamical potentials and provides a natural definition of order
parameters. The new definitions are constructed to be directly
operational from the experimental point of view. Finally, they show why
without the thermodynamic limit\index{thermodynamic limit} or at
phase-transitions, the systems do not have a single peaked
distribution in phase space.

In a closely related contribution, Dieter Gross~\cite{grossdd} makes
the statement, that the microcanonical ensemble with Boltzmann's
principle $S=k_B\ln\Omega$ is the only proper basis to describe the
equilibrium of a closed "small" system.  Phase-transitions are linked
to convex (upwards bending) intruders\index{convex intruder} of the
entropy, where the canonical ensemble defined by the Laplace transform
to the intensive variables becomes multi-modal, non-local, and
violates the basic conservation laws. The one-to-one mapping of the
Legendre transform being lost, Gross claims that it is all possible
without invoking the thermodynamic limit, extensivity, or concavity of
the entropy.

\section{Thermodynamics}

\subsection{Inequivalence of statistical ensembles}\index{statistical mechanics}

Following the example exhibited long time ago by Hertel and
Thirring~\cite{Thirringdd}, it is striking that these systems could lead
to inequivalences between microcanonical
\index{microcanonical!ensemble}, canonical 
\index{canonical!ensemble} or grand canonical
ensembles\index{grandcanonical ensemble}\index{ensemble inequivalence}. 
In this book, the first example is given by Barr{\'e}
{\it et al}~\cite{BMRdd} who present the Blume-Emery-Griffiths (BEG)
model which allows a deep understanding of the fundamental reason why
this happens. They studied the spin-1 BEG model both in the canonical
and in the microcanonical ensemble.  The canonical phase diagram
exhibits a first order and a continuous transition lines which join at
a tricritical point. It is shown that in the region where the
canonical transition is first order, the microcanonical ensemble
yields a phase diagram which differs from the canonical one. In
particular it is found that the microcanonical phase diagram exhibits
energy ranges with negative specific heat and temperature jumps at the
transition energies. The global phase diagrams in the two ensembles
and their multicritical behavior are calculated and compared.

\begin{figure}
\begin{center}
\includegraphics[width=.5\textwidth]{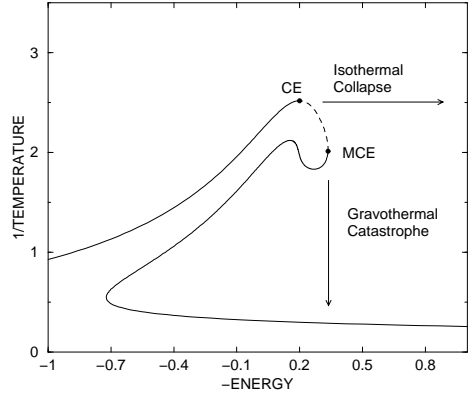}
\end{center}
\caption[]{Inverse temperature as a
function of the energy for self-gravitating fermions without
cut-off. CE and ME refers respectively to the
  tricritical points in the canonical and microcanonical ensembles.}
\label{chavanis}
\end{figure}

Pierre-Henri Chavanis~\cite{chavanis} shows similar features in
self-gravitating systems where canonical and microcanonical
tricritical points do not coincide either, as shown in
Fig.~\ref{chavanis} in the framework of self-gravitating fermions. Let
us emphasize that this property survives to the introduction of a
finite cut-off instead of quantum degeneracy as discussed by Chavanis.

\subsection{Negative specific heats}\index{negative!specific heat}

This fact produces striking phenomena in the microcanonical ensemble,
since it may result in a negative specific heat region, as was
emphasized~\cite{eddigton} by Eddington\index{astrophysics} in 1926
and then discussed by Lynden-Bell~\cite{Lynden-BelWood}.  A first
remark on the possibility of having negative specific heat in the
microcanonical ensemble can even be found in the seminal paper on
statistical mechanics by J.C. Maxwell~\cite{Maxwell}.
Thirring~\cite{Thirringdd} has finally clarified the point by showing
that the paradox disappears if one realizes that only the
microcanonical specific heat could be negative.

Indeed, in the canonical ensemble\index{canonical!ensemble}
 the mean value of the energy of
a system with different energy levels $E_i$ is
\begin{equation}
\langle E\rangle=\frac{\sum_i E_i\;e^{-\beta
E_i}}{Z}=-\frac{\partial \ln Z }{\partial \beta}
\end{equation}
where $Z$ is the partition function. It is then straightforward to
compute the specific heat
\begin{eqnarray}
{ C_v}=\frac{\partial \langle E\rangle}{\partial T}\propto\langle
\left(E   -\langle E\rangle\right)^2\rangle{ >0}\quad.
\end{eqnarray}
This clearly shows that the canonical specific heat is always
positive. Notice also that this condition is true for systems of any
size, regardless of whether a proper thermodynamic limit exists or not.

This is not the case if the energy is constant
\index{microcanonical!ensemble} as shows the simplified following derivation for the
example of interacting self-gravitating systems. Using the virial theorem for
 such particles
\begin{equation}
2\langle E_c\rangle+\langle E_{pot}\rangle=0\quad,
\end{equation}
  one gets that the total energy
\begin{equation}
  E=\langle E_c\rangle+\langle E_{pot}\rangle=-\langle E_c\rangle\quad.
\end{equation}
As the kinetic energy $E_c$ is by definition proportional of the
temperature one gets that
\begin{equation}
{C_v}=\frac{\partial E}{\partial T}\propto\frac{\partial
E}{\partial E_c} <0
\end{equation}
Loosing its energy, the system is becoming hotter.

\begin{figure}[ht]
\begin{center}
\includegraphics[width=.5\textwidth]{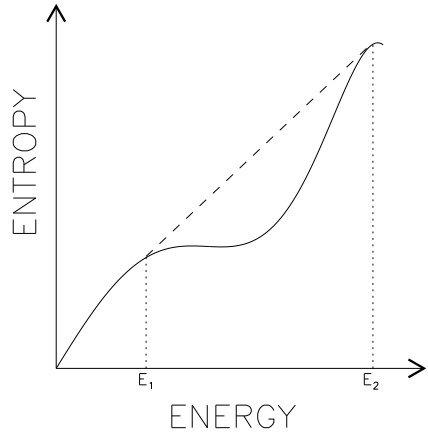}
\end{center}
\caption[]{Schematic shape of the entropy $S$ as a function of
the energy $E$ with a convex intruder: the solid curve corresponds
to the microcanonical result, whereas the dashed line to the
canonical one.} \label{concavite}
\end{figure}

It is important at this stage to make a short comment on the Maxwell
construction, usually taught in the framework of the Van der Waals
liquid-gas transition. The existence of a negative specific heat
region corresponds to a convex intruder\index{convex intruder}
\index{convexity} in the entropy-energy curve, as shown in
Fig.~\ref{concavite}. When the interactions are short range, the
system will phase separate in two parts, corresponding to the two
phases 1 and 2 with a molar fraction $x$, so that the free energy
$xF_1+(1-x)F_2$ is lower than the original free energy. This is
clearly possible if the energy cost of the interface is proportional
to the surface whereas the energy gain is proportional to the volume
of the phase. However, this is not any more possible when the
interactions are long range since, on one hand, it is not
straightforward to define a phase and, moreover, the system is not
additive.  The Maxwell construction has to be redefined in this new
framework.

\begin{figure}[ht]
\begin{center}

\includegraphics[width=.8\textwidth]{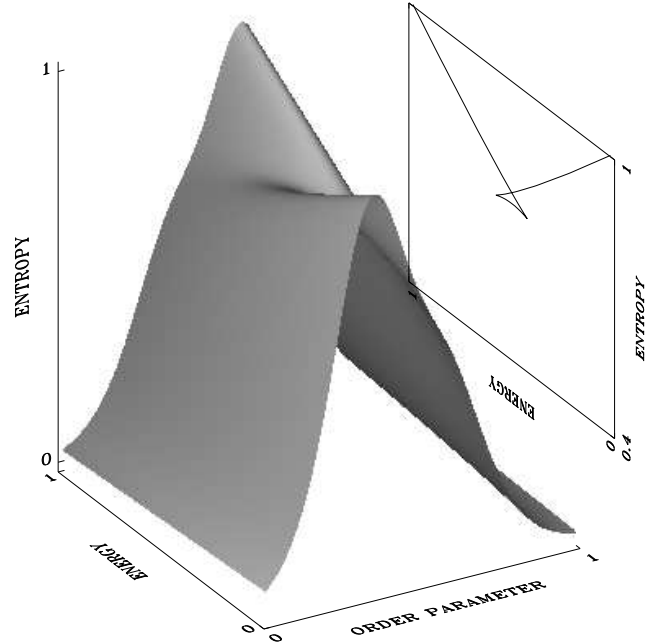}

\end{center}
\caption{A stylized microcanonical entropy as a function of the energy and of the
  order parameter mimicks an Alpine landscape where the workshop took
  place. The projection for the critical points of the surface onto
  the entropy-energy plane produces the well known "swallowtail"
  catastrophe.} \label{embleme}
\end{figure}

Let us note that the microcanonical entropy as a function of the
energy and of the order parameter generically leads to the landscape
presented in Fig.~\ref{embleme}. The projection for the critical
points of the surface onto the entropy-energy plane produces the well
known "swallowtail" catastrophe~\cite{PostonStewart}, depicted on the
right of the figure.

This concept of negative specific heat is now widely accepted in
the astrophysical community, and was popularized in particular by
Hawking~\cite{hawking} in 1974, with some esoteric applications to
black holes. The caloric curve of self-gravitating fermions
derived by Chavanis and shown in Fig.~\ref{chavanis} emphasizes
such negative specific heat branch: the dotted branch is one example.
Similarly one gets negative specific heat branch in the BEG model
proposed by Barr{\'e} {\it et al}~\cite{BMRdd}. In the canonical
ensemble, they correspond to local maxima or saddle point of the corresponding
free energy; it is the constraint of keeping the energy constant
that stabilizes these canonical unstable states in the
microcanonical ensemble\index{microcanonical!ensemble}.

Experimental groups have recently claimed signatures of negative
specific heats in small systems. The first one corresponds to nuclear
fragmentation~\cite{dagostino}\index{nuclear physics}, even if the
authors use prudently the word "indication" of negative specific heat.
The latter being inferred from the event by event study of energy
fluctuations from Au + Au collisions. However, the signatures
correspond to indirect measurements.

In the clusters\index{atomic!clusters} community, two experimental
groups have very recently reported negative specific heat. The first
system~\cite{haberland} correspond to atomic sodium clusters, namely
Na$_{147}^+$ and the negative microcanonical specific heat has been
found near the solid to liquid transition. The cluster ion are
produced in a gas aggregation source and then thermalized with Helium
gas of controlled temperature. Accelerated thanks to the charge in a
mass spectrometer, they are finally irradiated by a laser to determine
the energy from the evaporation of several atoms after laser
irradiation, also called photofragmentation.  However, the control of
equilibrium is as always the key point and therefore the evaluation of
the temperature seems to be questionable, in particular since the
temperature could not be constant during the motion of the ions.

In the Lyon's molecular cluster experiment~\cite{gobet}, with
H$_{17}^+$, the energy and the temperature are determined from the
size distribution of fragments after collision of the cluster with a
Helium projectile. To simplify the method, the larger the ratio of
small fragments versus large ones, the larger is the temperature
determined using the Bonasera et al procedure~\cite{bonasera}.  The
caloric curve reported~\cite{gobet} shows a plateau. Work along this
line is in progress and seems to show a negative specific heat
region.

\subsection{Non extensive statistics}\index{non!extensive thermodynamics}

Constantino Tsallis, Andrea Rapisarda, Vito Latora and Fulvio
Baldovin~\cite{RT} review the generalized non-extensive statistical
mechanics formalism and its implications for different physical
systems. The original very interesting idea is to generalize
Boltzmann's entropy by defining
\begin{eqnarray}
  S_{q}&=&k_B\ \frac{1-\sum_i{p_i}^{q}}{q-1}
\end{eqnarray}
where $\sum_ip_i=1$. Using either the L'Hopital rule or a first
order expansion of the term $p_i^{q}$ in power of $q$, one
immediately notices that
\begin{equation}
\displaystyle\lim_{q\to1}S_{ q}=-k_B\sum_ip_i\ln p_i\quad,
\end{equation}
i.e. the well known Shannon entropy, known to be equivalent to the
Boltzmann's one.

However, for $q$ different from 1, this generalized entropy
$S_{ q}$ is non additive, and one gets
\begin{eqnarray}
S_{ q}(A+B)=S_{ q}(A)+S_{ q}(B)+\left(1-{ q}\right)S_{ q}(A)S_{
q}(B)\quad.
\end{eqnarray}
They illustrate in particular its application and the meaning of
the entropic index~$q$ for conservative and dissipative
low-dimensional maps. They also report on non Boltzmann-Gibbs
behavior~\cite{RT}  and hindrance of relaxation for Hamiltonian systems with
long-range interaction, where fingerprints of the generalized
statistics have recently emerged.

This very interesting proposal~\cite{Tsallisdd} had however until now
no strong foundations and many physicists were not ready to admit that
the exponential Boltzmann distribution of states is at equilibrium
only a particular case of a generalized distributions, with power
tails.  Dieter Gross~\cite{grossdd} in particular makes different
comments to this point. On the contrary, Tsallis et al emphasize also
different situations were the Boltmann-Gibbs behavior is clearly not
appropriate. 

Recently, Beck and Cohen~\cite{beckcohen} showed that considering
different statistics with large fluctuations, one can obtain
generalized results, called superstatistics, with the Tsallis
formalism being presumably so far the most relevant example.
Moreover, Baldovin and Robledo worked out~\cite{baldoreblo,RT} exactly
the $q$ indices for the generalized largest Lyapunov exponent proposed
by Tsallis. This an important step toward the derivation of a complete
theory which, in particular, should help to understand the limits of
its applications.

\section{Dynamical aspects}

An essential peculiarity of these physical systems, and of some of
their simplified models, is that a classical system of particles
with long range interactions will display strong non-equilibrium
features. Dynamics is typically chaotic and self-consistent, since
all particles give a contribution to the field acting on each of
them: one calls this {\em self-consistent chaos}. Numerous
physical systems fall in this category: galactic dynamics,
dynamics of a plasma, vorticity dynamics,....

It is therefore essential to study the thermodynamic stability of
these systems and in particular to understand the formation of
structures\index{coherent!structures} trough {\em instabilities}.
They should have logical similarities with the Jean's instability of
self-gravitating systems, or with the modulational instability,
leading to the formations of localized structures, as confirmed by
preliminary results. Additional dynamical effects, like anomalous
diffusion\index{anomalous!diffusion} and Levy walks, which are
reported in the negative specific heat regions, should be linked to
these uncommon characteristics of thermodynamics~\cite{Torcini}.

In particular, Diego Del Castillo Negrete~\cite{diego} discusses a
mean-field single-wave model\index{mean field!models} that describes
the collective dynamics of marginally stable fluids and
plasmas\index{plasma!physics}. He shows thus the role of
self-consistent chaos\index{self!consistent chaos} in the formation
and destruction of coherent structures, and presents a mechanism for
violent relaxation of far from equilibrium initial
conditions\index{non!equilibrium phenomena}. The model bears many
similarities with toy-models used in the study of systems with long
range interactions in statistical mechanics, globally coupled
oscillators, and gravitational systems.

One of these toy models is for example studied by Dauxois et
al~\cite{DLRRT}. They consider the dynamics of the Hamiltonian Mean
Field model which displays several interesting and new features. They
show in particular the emergence of collective properties, i.e. the
coherent (self-consistent) behavior of the particle motion. The
space-time evolution of such coherent structures can sometimes be
understood using the tools of statistical mechanics
\index{statistical mechanics}, otherwise can be a manifestation of 
the solutions of an
associated Vlasov equation. Both cases in which the interaction among
the particles is attractive and the one where it is repulsive are
interesting to study: they offer different views to the process of
cluster formation and to the development of the collective motion on
different time-scales. The clustering transition can be first or
second order, in the usual thermodynamical sense. In the former case,
ensemble inequivalence\index{ensemble inequivalence} naturally arises
close to the transition.  The behavior of the Lyapunov spectrum is
also commented and the 'universal' features of the scaling laws that
it involves.

Yves Elskens~\cite{elskens} shows that plasmas\index{plasma!physics}
are a most common example of systems with long-range interactions,
where the interplay between collective (wave) and individual
(particle) degrees of freedom is well known to be central. This
interplay being essentially non-dissipative, its prototype is
described by a self-consistent Hamiltonian, which provides clear and
intuitive pictures of fundamental processes such as the weak warm beam
instability and Landau damping in their linear regimes. The
description of the nonlinear regimes is more difficult. In the damping
case, new insight is provided by a statistical mechanics approach,
which identified the distinction between a trapping behavior and
linear Landau behavior in terms of a phase transition. In the unstable
case, the model has shown that the commutation of long-time and
large-$N$ limits\index{thermodynamic limit} is not guaranteed.

Chavanis considers also dynamical aspects in the framework of stellar
systems and two-dimensional vortices. He discusses in particular
two possible relaxation scenari: one due to collisions (or
more precisely to discrete interactions) and the second one,
called violent relaxation, really collisionless but due to the mean
field effect and the long range of the interaction.

Finally, the dynamical processes that give rise to power-law
distributions and fractal structures have been studied extensively in
recent years. Ofer Biham and Ofer Malcai~\cite{biham} describe recent
studies of self-organized criticality in sandpile models as well as
studies of multiplicative dynamics, giving rise to power-law
distributions.  Sandpile models turn out to exhibit universal behavior
while in the multiplicative models the powers vary continuously as a
function of the parameters. They consider the formation of a fractal
object in the presence of a dynamical mechanism that generates a
power-law distribution and presents a model that demonstrates
clustering when the probability of adding a particle decays with a
power $\alpha > d$, so it has a short-range nature.

\section{Bose-Einstein condensation}\index{Bose!Einstein condensation}

Finally, we would like to put a special emphasis on Bose Einstein
Condensation (BEC), predicted by Bose and Einstein in 1924, which
could be an important field of applications. With the recent
achievement~\cite{expBEC} of Bose-Einstein condensation in atomic
gases thanks to the evaporation cooling technique, it becomes possible
to study these phenomena in an extremely diluted fluid, thus
helping to bridge the gap between theoretical studies, only
tractable in dilute systems, and experiments. In the BEC, atoms
are trapped at such low temperatures that they tumble into the
same quantum ground state creating an intriguing laboratory for
testing our understanding of basic quantum phenomena.

First, Jean Dalibard~\cite{dalibard} presents how coherence and
superfluidity are hallmark properties of quantum fluids and
encompass a whole class of fundamental phenomena. He reviews
several experimental facts which reveal these two remarkable
properties. Coherence appears in interference experiments, carried
out either with a single condensate or with several condensates
prepared independently. Superfluidity can be revealed by studying
the response of the fluid to a rotating perturbation, which
involves the nucleation of quantized vortices.

Second, Ennio Arimondo and Oliver Morsch~\cite{arimondo} present the
current investigations of Bose-Einstein condensates  within optical
lattices, where the long range interactions are an essential part of
the condensate stability. Previous work with laser cooled atomic
gases is also briefly discussed.

On the theoretical side, the fluctuations of the number of particles
in ideal Bose-Einstein condensates within the different statistical
ensembles has shown interesting differences. Martin Holthaus
explains~\cite{holthauss} why the usually taught grand canonical
ensemble\index{grandcanonical ensemble} is inappropriate for
determining the fluctuation of the ground-state occupation number of a
partially condensed ideal Bose gas: it predicts
r.m.s.-fluctuations\index{anomalous!condensate fluctuations} that are
proportional to the total particle number at vanishing temperature. In
contrast, both the canonical\index{canonical!ensemble} and the
microcanonical ensemble yields fluctuations that vanish properly for
the temperature going toward zero. It turns out that the difference
between canonical and microcanonical fluctuations can be understood in
close analogy to the familiar difference between the heat capacities
at constant pressure and at constant volume.  The detailed analysis of
ideal Bose-Einstein condensates turns out to be very helpful for
understanding the occupation number statistics of weakly interacting
condensates.

Ulf Leohnardt~\cite{ulf} shows that Bose-Einstein condensates can
serve as laboratory systems for tabletop astrophysics. In particular,
artificial black holes can be made (sonic or optical black holes). A
black hole represents a quantum catastrophe where an initial
catastrophic event, for example the collapse of the hole, triggers a
continuous emission of quantum radiation (Hawking radiation). The
contribution summarizes three classes of quantum catastrophes, two
known ones (black hole, Schwinger's pair creation) and a third new
class that can be generated with slow light.

Finally, Gershon Kurizki presents~\cite{gershon} an exciting
theoretical idea to induce long-range attractions between atoms that
acts across the whole Bose-Einstein condensate. He shows that
dipole-dipole interatomic forces\index{dipolar interaction} induced by
off-resonant lasers
\begin{eqnarray}
V_{dd}=V_0\left[\frac{2z^2-x^2-y^2}{r^{3}}(\cos qr+qr\sin
qr)-\frac{2z^2+x^2+y^2}{r}q^2\cos qr \right]
\end{eqnarray}
allow controllable drastic modifications of cold atomic media.
"Sacrifying strength for beauty", Kurizki proposed~\cite{Kurizki} to average out
the first term in $1/r^3$ of the dipole-dipole interaction by the
different lasers, in order to keep only the last one with a $1/r$
interaction. The important point is that induced gravity-like\index{gravitation}
force would be strong enough to see it acting among atoms in the
BEC: i.e. that, having induced the gravity-like attraction in the
BEC, one could switch off the trap used originally to create the
BEC, and it will remain stable, holding itself together.

Depending on the number of lasers, the resulting gravity-like force
could be aniso\-tro\-pic for three lasers, or strictly identical
to gravity with eighteen lasers~! If the last proposal is
presumably too
speculative and if the difficulties (the power of the laser
required being really huge) facing the experimentalists
are a real challenge, the ability to emulate gravitational
interactions in the laboratory is of course fascinating. Indeed,
these modifications may include the formation of self-gravitating
"bosons stars" and their plasma-like oscillations, self-bound
quasi-onedimensional Bose condensates  and their "supersolid"
density modulation, giant Cooper pairs and quasibound molecules in
optical lattices  and anomalous scattering spectra in  systems of
interacting Bosons or Fermions. These novel regimes  set the arena
for the exploration of exotic astrophysical and condensed -matter
objects, by studying their atomic analogs {\em in the laboratory}.

\section{Conclusion}
The dynamics and thermodynamics of long range system is a rich and
fascinating topic in particular for the following issues:
\begin{itemize}
\item Statistical physics\index{statistical mechanics}: inequivalence
  of ensembles, negative specific heat, collisionless relaxation, role
  of coherent structures, relationship between dynamics and
  thermodynamics, nonadditivity, generalization of the entropy,...

\item This problem has also the nice property to be related to
  neighbooring scientific disciplines. Not only mathematics with
  application of catastrophe theory~\cite{barrebouchet}, large
  deviations theory~\cite{ellisdd}, but also to computer science;
  because of the long range interactions naive numerical code are of
  order $N^2$, and need the developments of efficient algorithms such
  as the heap based procedure~\cite{heap}, local simulation algorithm
  for Coulomb interaction~\cite{maggs},...

\item However, this methodological and fundamental effort should
  provide a general approach to the problems arising in each specific
  domain which has motivated this study: astrophysical objects,
  plasmas, atomic and molecular clusters, fluid dynamics, fracture,
  Bose-Einstein condensation, ... in order to detect the depth and the
  origin of the observed analogies or, on the contrary, to emphasize
  their specificities.

\end{itemize}

Many of these different aspects are considered in this book but it
is clear that, rather than closing the topic, it opens the pandora
box.

\section*{Acknowledgements}
This work has been partially supported by th EU contract No.
HPRN-CT-1999-00163 (LOCNET network), the French Minist{\`e}re de la
Recherche grant ACI jeune chercheur-2001 N$^\circ$ 21-311.  This work
is also part of the contract COFIN00 on {\it Chaos and localization in
  classical and quantum mechanics}.


\begin{thebibliography}{8.}
\addcontentsline{toc}{section}{References}

\bibitem{Padmanabhan} T. Padmanabhan: Physics Reports \textbf{188}, 285   (1990)

\bibitem{KacUhlenbeck} M.~Kac, G.E.~Uhlenbeck, P.C.~Hemmer: J. 
Math. Phys. \textbf{4},
   216 (1963)

\bibitem{leshouches} T. Dauxois, S. Ruffo, E. Arimondo, M. Wilkens:
``Dynamics and Thermodynamics of Systems with Long Range
Interactions'', Lecture Notes in Physics Vol. 602, Springer (2002)
(this volume).

\bibitem{BMRdd} J. Barr{\'e}, D. Mukamel, S. Ruffo: \emph{Inequivalence of 
ensembles in mean-field models of magnetism} in~\cite{leshouches} (in this volume)

\bibitem{paddy} T. Padmanabhan: \emph{Statistical mechanics of
     gravitating systems in static and expanding backgrounds}
   in~\cite{leshouches} (in this volume)

\bibitem{chavanis} P.-H. Chavanis:
\emph{Statistical mechanics of two-dimensional vortices and
 stellar systems} in~\cite{leshouches} (in this volume)

\bibitem{cohen} E. G. D.  Cohen, I. Ispolatov:
\emph{Phase transitions in systems
with $1/r^\alpha$ attractive interactions} in~\cite{leshouches} (in this volume)


\bibitem{elskens} Y. Elskens:
\emph{Kinetic theory for plasmas and wave-particle hamiltonian dynamics} in~\cite{leshouches} (in this volume)

\bibitem{diego} D. Del Castillo-Negrete:
\emph{Dynamics and self-consistent chaos  in a mean field
 Hamiltonian model} in~\cite{leshouches} (in this volume)

\bibitem{muskhelishvili}
N. I. Muskhelishvili, in \emph{Some Basic Problems of the Mathematical 
  Theory of Elasticity}, P. Noordhoff, Groningen (1953)

\bibitem{Landau} L. D. Landau, E. M. Lifshitz: \emph{Course of
Theoretical Physics. T. 8: Electrodynamics of continuous media} (1984)

\bibitem{Griffiths} R. B. Griffiths: Physical Review \textbf{176}, 655 
  (1968). M.E. Fisher, Arch. Rat. Mech. Anal. \textbf{17}, 377 
  (1964)


\bibitem{chomazdd} Ph. Chomaz, F. Gulminelli:
\emph{Phase transitions in finite systems} in~\cite{leshouches} (in this volume)

\bibitem{Dieter} D. H. E. Gross: \emph{Microcanonical Thermodynamics},
  World Scientific, Singapore (2001)

\bibitem{grossdd} D. H. E. Gross:
\emph{Thermo-Statistics or Topology of the Microcanonical Entropy
  Surface} in~\cite{leshouches} (in this volume)

\bibitem{Thirringdd} W. Thirring: Z. Phys. \textbf{235}, 339 (1970)

\bibitem{eddigton}
A. S. Eddington: \emph{The internal constitution of stars},
Cambridge University Press (1926)

\bibitem{Lynden-BelWood} D. Lynden-Bell, R. Wood: Mont. Not. R.
   Astron. Soc. \textbf{138}, 495 (1968)

\bibitem{Maxwell} J.C. Maxwell: {\it On Boltzmann's theorem on the
average distribution of energy in a system of material points}, Cambridge
Philosophical Society's Trans., vol XII, p. 90 (1876)

\bibitem{PostonStewart}
T. Poston, J. Stewart, \emph{Catastrophe Theory and its Application},
Pitman, London (1978)

\bibitem{hawking} S. W. Hawking: Nature \textbf{248}, 30 (1974)

\bibitem{dagostino}
D'Agostino et al: Physics Letters B \textbf{473}, 219 (2000)

\bibitem{haberland}
M. Schmidt et al: Physical Review Letters \textbf{86}, 1191
(2001)

\bibitem{gobet}
F. Gobet et al: Physical Review Letters \textbf{87}, 203401
(2001)

\bibitem{bonasera}
M. Belkacem, V. Latora, A. Bonasera: Physical Review C
\textbf{52}, 271 (1995)

\bibitem{RT} C. Tsallis, A. Rapisarda, V. Latora, F.  Baldovin:
\emph{Nonextensivity: from low-dimensional maps to Hamiltonian
systems} in~\cite{leshouches}  (in this volume)

\bibitem{Tsallisdd} C.  Tsallis: Journal of Statistical Physics
   \textbf{52}, 479 (1988)

\bibitem{beckcohen} C. Beck, E.G.D. Cohen, [cond-mat/0205097]

\bibitem{baldoreblo} 
F. Baldovin and A. Robledo, [cond-mat/0205356].

   
\bibitem{Torcini} A. Torcini, M. Antoni: Physical Review E
   \textbf{59}, 2746 (1999)

\bibitem{DLRRT} T. Dauxois, V. Latora, A. Rapisarda, 
S. Ruffo, A.
   Torcini: \emph{The Hamiltonian mean field model: from dynamics to
     statistical mechanics and back} in~\cite{leshouches} (in this volume)

\bibitem{biham} O. Biham, O. Malcai: \emph{Fractals and Power-Laws} in~\cite{leshouches} (in this volume)

\bibitem{expBEC}
M. H. Anderson et al: Science \textbf{269}, 198 (1995). C. C.
Bradley et al, Physical Review Letters \textbf{75}, 1687 (1995).
K. B. Davis et al, Physical Review Letters \textbf{75}, 3669
(1995)

\bibitem{dalibard} J. Dalibard:
\emph{Coherence and superfluidity of gaseous Bose-Einstein
condensates} in~\cite{leshouches} (in this volume)

\bibitem{arimondo} O. Morsch, E. Arimondo:
\emph{Ultracold atoms and Bose-Einstein condensates in optical
   lattices} in~\cite{leshouches} (in this volume)

\bibitem{holthauss} D. Boers, M. Holthaus: \emph{Canonical statistics of
    occupation numbers for ideal and weakly interacting Bose-Einstein
    condensates} in~\cite{leshouches} (in this volume)

\bibitem{ulf} U. Leonhardt: \emph{ Quantum catastrophes}
in Proceedings of the Conference "Dynamics and thermodynamics of
systems with long range interactions", Les Houches, France, February
18-22 2002,   T. Dauxois, E. Arimondo, S. Ruffo, M. Wilkens Eds.,
published on http://www.ens-lyon.fr/$\sim$tdauxois/procs02/

\bibitem{gershon} G. Kurizki, D. O'Dell,
S. Giovanazzi,
A. I. Artemiev: \emph{New regimes in cold gases via
   laser-induced long-range interactions} in~\cite{leshouches} (in this volume)

\bibitem{barrebouchet} J. Barr{\'e}, F. Bouchet, in preparation
(2002)

\bibitem{Kurizki} D. O'Dell, S. Giovanazzi, G. Kurizki, V. M.
   Akulin: Physical Review Letters \textbf{84}, 5687 (2000)

\bibitem{ellisdd} R. S. Ellis, K. Haven, B. Turkington: \emph{The large
    deviation principle and complete equivalence and nonequivalence
    results for pure and mixed ensembles}, Journal of Statistical
  Physics \textbf{101}, 999 (2000)

\bibitem{heap} A. Noullez, D. Fanelli, E. Aurell:
``Heap base algorithm'', cond-mat/0101336 (2001)

\bibitem{maggs} A. C. Maggs, V. Rossetto, Physical Review Letters
  \textbf{88}, 196402 (2002)

\end{thebibliography}
\end{document}